\documentstyle[aps,prl,twocolumn,epsfig,amsmath,amssymb]{revtex}
\begin{document}

\twocolumn[\hsize\textwidth
\columnwidth\hsize\csname
@twocolumnfalse\endcsname

\title{Microscopic kinetics and time-dependent structure factors}

\author{T.~Aspelmeier, B.~Schmittmann, and R.~K.~P.~Zia}
\address{Center for Stochastic Processes in Science and Engineering and\\
Department of Physics, Virginia Tech, Blacksburg, VA 24061-0435 USA}
\date{January 12, 2001}

\maketitle

\begin{abstract}
  The time evolution of structure factors in the disordering process of an
  initially phase separated lattice depends crucially on the microscopic
  mechanism which drives the disordering, such as Kawasaki dynamics or vacancy
  mediated disordering.  Monte Carlo (MC) simulations show unexpected ``dips''
  in the structure factors, which mean-field theory completely fails to
  capture.  The disordering via vacancies is slower by a surprisingly large
  constant factor compared to Kawasaki dynamics. To understand the dips in odd
  structure factors (i.e., those reflecting the initial order), a
  phenomenological model is introduced, and an analytical solution of Kawasaki
  dynamics is derived, in excellent agreement with simulations. The presence
  of dips in the even structure factors for vacancy mediated, and their
  absence for Kawasaki dynamics, marks a significant, but not yet understood,
  difference of the two dynamics.
\end{abstract}
\vspace{2mm} 
]

\narrowtext

After a temperature ``downquench'' into a coexistence region, binary systems
such as magnets or alloys develop long-range order in a universal fashion 
\cite{exp}. Determined by symmetries and conservation laws, the dynamic
coarsening follows one of a small number of scenarios, characterized by
dynamic scaling and universal exponents. Considerable work has been devoted
to the theoretical foundations of this behavior \cite{th}. By contrast,
far less attention has focused on the opposite phenomenon, i.e.\ {\it %
disordering} following an {\it upquench} in temperature, 
even though it is hardly less important in materials science:
It determines, e.g., the time evolution of an interface between two
materials, or the waiting times which must elapse before a heated
material has mixed sufficiently.

A key ingredient in the physics of these processes is the kinetic mechanism
by which two particles exchange. In many materials, it is
dominated by the diffusion of vacancies or defects \cite{VMD}, rather than
direct atom-atom exchanges. Vacancy concentrations are typically very small (%
$\sim $ $10^{-5}$) \cite{MS}. For this type of dynamics, it has recently
been shown \cite{MFT} that a simple mean-field theory suffices to obtain a 
{\em quantitative} description of disordering characteristics which are
controlled by short-range correlations. 

In this letter, we will show that this mean-field theory fails dramatically,
even at the qualitative level, to predict the dynamics of observables which
include correlations over {\em larger} distances in the system. A typical
example is the time-dependent equal-time structure factor.
We present MC simulation data and a
microscopic theory, illustrating the crucial role played by the details of
the whole vacancy path. Comparing data for vacancy-mediated and direct
exchange dynamics, we obtain a key result of our study, namely, that
experimental data for the equal-time structure factor can identify the {\em %
microscopic kinetic mechanism} by which a material evolves. We conclude with
a list of open questions.

We consider a $d$-dimensional square lattice of size $L^{d}$ ($L$ {\em even},
for simplicity) with periodic boundary conditions. Each site $\vec{\imath}$ is
occupied by one of two types of particles, which we label as spins: $s_{\vec{%
\imath}}=\pm 1$. A single spin is {\em tagged}, playing the role of a vacancy
or defect. Only this ``vacancy spin'' is allowed to move, all others being
passive. Turning to the microscopic dynamics, we define two
versions: (\textit{i}) The vacancy performs a Brownian random walk on the
lattice, exchanging with nearest-neighbor spins and thus scrambling the spin
configuration, or (\textit{ii}) the tagged spin performs one single step, and
then passes the tag to another, randomly selected spin. This sequence is then
repeated. Note that both versions strictly conserve particle number (or total
magnetization, $\sum_{\vec{\imath}}$ $s_{\vec{\imath}}$). The first
is known as vacancy mediated dynamics (VMD) \cite{VMD,MS}, the second 
is precisely Kawasaki dynamics (KD) \cite{KD}. In the language of Ising
models, they correspond to $T=\infty $ dynamics, since energy costs are
ignored. Yet, we consider a perfectly phase-segregated initial
configuration (up- and down-spins separated by two planar domain walls
perpendicular to, say, the $x$-axis), corresponding to $T=0$. At time $t=0$,
the system experiences a temperature upquench, from zero to infinity, and we
monitor the disordering process as a function of time. For $t\rightarrow
\infty $, the system clearly reaches the infinite temperature equilibrium
state, characterized by completely random configurations. 
%Our focus is the time dependence of $\left\{ s_{\vec{\imath}}(t)\right\} $, as
%the system evolves towards equilibrium.

We monitor the time-dependent structure factors,
$S_{\vec{k}}(t)\equiv L^{-d}\left\langle \left| \sum_{\vec{\jmath}}
s_{\vec{\jmath}}(t)\exp (-i \vec{k}\cdot
\vec{\jmath})\right|^{2}\right\rangle$. Here, $\left\langle \bullet
\right\rangle$ denotes the time-dependent average over a large number ($1,000$
for our data) of independent runs, starting from the same $T=0$
configuration differing only in the random initial position of the 
tag. MC time is incremented by $1$ for each exchange. The most
interesting structure factors are those whose wave vector $\vec{k}$ is
perpendicular to the initial phase boundary, $\vec{k}=(2\pi n/L,0,\dots ,0)$,
with $n$ integer. For odd $n$ $\ll L$, they are sensitive to the initial order
so that $S_{\vec{k}}(0)$ is of the order $L^{d}$, while for even $n$,
$S_{\vec{k}}(0)=0$ (since $L$ is even). For brevity, these two classes will be
referred to as odd and even $\vec{k}$. Clearly, 
$\lim_{t\rightarrow \infty \text{ }}S_{\vec{k}}(t)=1$ for all $\vec{k}$, due
to the random final state.

\begin{figure}[tb]
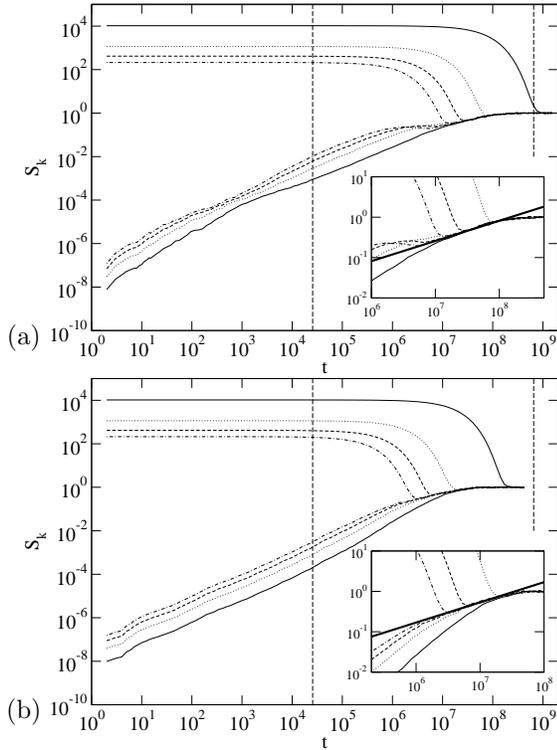

\centerline{\makebox[0cm]{\raisebox{.4cm}{(a)}}\epsfig{file=fig1a,width=2.8in}}
\centerline{\makebox[0cm]{\raisebox{.4cm}{(b)}}\epsfig{file=fig2a,width=2.8in}}
\caption{Time evolution of the first eight structure factors $S_{\vec{k}}$
with $\vec{k}=(2\pi n/L,0)$, $n=1,3,5,7,8,6,4,2$ from top to bottom, on a
$d=2$ lattice of size $L=160$ with (a) VMD and (b) KD. The thick
solid straight lines in the insets, which show the magnified crossover region,
have slope $1/2$. The two dashed vertical lines window the intermediate
regime.}
\label{fig:1}
\end{figure}

For reference, we review a few key results from \cite{MFT}. The
disordering process displays three regimes, separated by two characteristic
times, $t_E\simeq L^2$, and $t_L\simeq L^{d+2}$. For $t\lesssim t_E$ only a
fraction of sites has been visited by the vacancy, and the 
equilibrium state is reached for $t\gtrsim t_L$. Between these
well-separated bounds lies the ``intermediate regime,'' in which the vacancy
visits each site many times without destroying much of the initial order. Here, 
the disorder parameter, being the number of broken nearest neighbor bonds (NN),
displays dynamic scaling with universal exponents before reaching its equilibrium
value.

Turning to structure factors, in Fig.~\ref{fig:1} we compare MC
data for a $160\times 160$ lattice, with (a) VMD and (b) KD. First, note
that three regimes emerge again. The final regime is easily identified 
by all $S_{\vec{k}}$'s reaching unity. In the intermediate
regime, the odd $S_{\vec{k}}$'s clearly display the decay of the initial
order which remained largely unaffected during the earliest regime. Second,
we emphasize a rather peculiar feature, namely, that several structure
factors develop a minimum, manifested as a small ``dip'' before they finally
tend to their equilibrium value. For VMD this dip appears
in {\em all} structure factors {\em except} the two lowest ones, i.e., $n=1,2$
(strictly speaking, in $n=4$ it emerges as a ``shoulder''). 
In contrast, for KD\ it exists only in the {\em odd} $S_{\vec{k}}$'s.
At first sight, these dips might suggest a temporary rebound
towards {\em order}. We will show below, however, that their origin is 
different.

\begin{figure}[tb]
\centerline{\epsfig{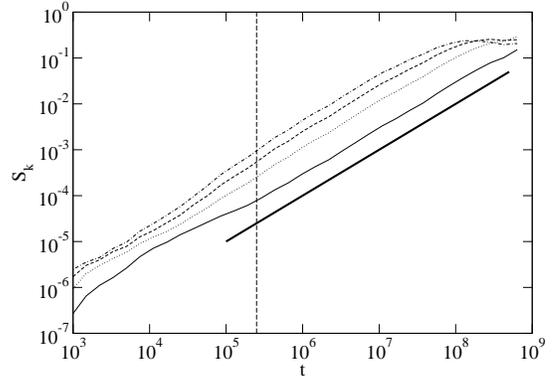}}
\caption{Short time behavior of the first 4 even structure factors on a $d=2$%
, $L=500$ lattice with VMD, averaged over $1,000$ runs. The thick solid line
has slope 1, the vertical line indicates the beginning of the intermediate
region.
% We also point out the onset of the dips at later times.
}
\label{fig:4}
\end{figure}

Third, turning to ``short'' times, just after the onset of the intermediate
regime, we observe a clear difference in the {\em even} structure factors for
the two types of dynamics, displayed more visibly in the data for a much
larger ($L=500$) lattice (Fig. \ref{fig:4}): The short-time behavior for a VMD
system appears to be $S_{\vec{k}}\sim t^{\gamma _{\text{V}}}$ with a
short-time exponent $\gamma _{\text{V}}\approx 1$, whereas the KD system
follows $S_{\vec{k}}\sim t^{\gamma _{\text{K}}}$ with $\gamma
_{\text{K}}>1$. Below, we will see that $\gamma _{\text{K}}=3/2$ exactly, in
the $L\rightarrow \infty $ limit. Thus, the presence or absence of dips
and the short-time behavior offer a simple signature to distinguish VMD from
KD. Finally, Fig.~\ref{fig:1} indicates that disordering by KD
is faster than VMD by a factor of about 4.5. This factor is roughly
independent of the lattice size but decreases with $d$ \cite{ASZ_long}.
We emphasize that, in all figures, the error bars are of the order of the
line thickness and therefore not shown.

We now turn to the theoretical descriptions of our model. Our first
approach, an intuitive model of two ``fluids,'' is based on the
coarse-grained mean-field theory of \cite{MFT} and accounts for the dips in
the odd structure factors. The second method involves an exact equation of
motion for the microscopic spins $s_{\vec{\imath}}(t)$, which can be solved
for the KD case and leads to $\gamma _{\text{K}}=3/2$.

\paragraph*{A ``two-fluid'' model.}

To motivate this approach, we briefly recall the key ingredients of the
mean-field theory \cite{MFT}. Coarse-graining the microscopic dynamics in
space and time, we identify two slow variables, namely, the local vacancy
and magnetization densities, $\varphi (\vec{r},t)$ and $\psi (\vec{r},t)$.
In the intermediate regime, the vacancy distribution has already reached its
equilibrium value, $\varphi _{o}=L^{-d}$, so that the only non-trivial
dynamics is embodied in $\psi (\vec{r},t)$, which obeys a simple diffusion
equation: $\partial _{t}\psi =D\nabla ^{2}\psi $. Here, the diffusion
coefficient $D\propto \varphi _{o}$ reflects the presence of a single
vacancy. The solution, with initial condition $\psi (\vec{r},0)=$ ${\rm sgn}%
(x)$, $|x|\leq L/2$, and periodic boundary conditions, is $\psi
(x,t)=\sum_{k}\frac{4}{ikL}{\rm e}^{ikx}{\rm e}^{-Dk^{2}t}\delta _{n\text{
odd}}$, where $k=2\pi n/L$ and $\delta _{n\text{ odd}} \equiv 1$ for $n=\pm 1,\pm
3,\dots$ and $0$ otherwise.

We propose to describe the local spin density by the {\em incoherent sum} of
two components (``fluids''), representing the ordered and disordered
fractions. The {\em ordered} component, starting from being 100\% phase
segregated and ending at zero, can be taken as $\psi (x,t)$. The remaining
component, with density $1-|\psi (x,t)|$, is {\em fluctuating} so that it will
be represented by $(1-|\psi (x,t)|)\eta (x,t)$, where $\eta (x,t)$ is a
delta-correlated noise with zero mean. Summarizing, we write the{\em \ spin
density} $s(x,t)$ as
\begin{equation}
s(x,t)=\psi (x,t)+(1-|\psi (x,t)|)\eta (x,t),  \label{eq:twofluid}
\end{equation}
Taking the Fourier transform of $s(x,t)$ and averaging over $\eta $ to
obtain $S_{k}(t)$, we find the final result
\begin{equation}
S_{k}(t)=L^{d}\left[ \frac{2e^{-n^{2}t/\tau }}{\pi n}\right] ^{2}\delta _{n%
\text{ odd}}+\sum_{n'\text{ odd}}\left[ 2\frac{%
1-e^{-n^{\prime 2}t/\tau }}{\pi n^{\prime }}\right] ^{2}.  \label{eq:S(k)}
\end{equation}
where $\tau \equiv L^{2}/(4\pi ^{2}D)$ plays the role of $t_{L}$. The
first term captures the decay of the initial ordered fraction. Note that only
the {\em odd} $k$'s are present. By contrast, the second term is $k${\em
-independent} and reflects the buildup of the disordered component.

The origin of the dips is now apparent: The first term starts with a
macroscopic amplitude, but decays on the time scale $\tau /2n^{2}$. Meanwhile,
due to the sum over all $n^{\prime }$, the second term rises to unity with
$\tau$, regardless of $n$. Therefore, $S_{k}(t)$, for odd $k$, does not
decay to unity monotonically. Further, these dips occur at earlier times for
larger $k$'s, a feature which agrees with the data, at least
qualitatively (Fig.~\ref{fig:1}). We believe that the contrasting terms may be
interpreted physically, the first being associated with the relatively {\em fast}
deterioration of the initial phase boundary while the second represents the
{\em slower} buildup of the disordered component.
%%%

Eq.~(\ref{eq:S(k)}) leads us to another prediction, namely, how the final
state is approached. By Poisson resummation \cite{Poisson}, it can be shown
that the second term increases as $t^{1/2}$ for times $t\lesssim \tau $.
This effect is most easily seen in the even $S_{\vec{k}}$'s where the
ordered contribution is absent. The simulation data show excellent agreement
with this behavior for both VMD and KD (Fig.~\ref{fig:1}).

Not surprisingly, this simple approximation has its shortcomings, such as $%
\sum_kS_k(t)\neq L^d$. Also, according to Eq.~(\ref{eq:S(k)}), all even
structure factors share the same behavior, which is clearly not borne out by
the simulations. Moreover, Eq.~(\ref{eq:S(k)}) gives no indication of an
early-time crossover from a $t^\gamma $ to a $t^{1/2}$ behavior. 
To improve our understanding, we now turn to an {\em exact} formulation of
the dynamics.

\paragraph*{Exact lattice calculation.}

In the following, we analyze the microscopic motion of the vacancy on the
lattice. The resulting equation of motion for the structure factor is, of
course, exact. Remarkably, it can be solved in the case of KD, providing us
with detailed information about different temporal regimes and finite size
effects. Here, we summarize the key features of this approach, while
deferring all details to a later publication\cite{ASZ_long}.

%Since our problem is equivalent to a non-interacting Ising model, 
Clearly, the
dynamics at each time step (for both VMD and KD) depends on the exchange of 
{\em only} {\em two} spins, so that the Liouvillian in the
master equation $\partial _{t}P(\{s_{\vec{\imath}}\};t)=\sum_{\{s_{\vec{%
\imath}}^{\prime }\}} \mathcal{L}$
$(\{s_{\vec{\imath}}\};\{s_{\vec{\imath}}^{\prime}\})
P(\{s_{\vec{\imath}}^{\prime }\};t)$
consists of a sum over terms of the
form $\delta (s_{\vec{\imath}}^{\prime };s_{\vec{\jmath}})\delta (s_{\vec{%
\imath}};s_{\vec{\jmath}}^{\prime })$. As a result, the equation for $G(\vec{%
x};t)$, the two-point function, is {\em linear} and involves no higher
correlations. Specifically, in the KD case, one obtains
(for $\vec{x}\neq \vec{0}$)
\begin{equation}
\label{eq:G}
\frac{dL^{d}}{2} \partial _{t}G(\vec{x};t)=\Delta _{\vec{x}}G(\vec{x}%
;t)+\sum_{\{\vec{b}\}}\delta_{\vec{x},\vec{b}} \left[ G(\vec{b}%
;t)-G(\vec{0};t)\right] 
\end{equation}
where $\partial _{t}$, $\Delta _{\vec{x}}$, and $\{\vec{b}\}$ stand for discrete time
differences, the lattice Laplacian and the set
of $2d$ lattice vectors, respectively. Of course, we may replace $G(%
\vec{0};t)$ by unity. Taking the Fourier transform of Eq.~(\ref{eq:G}), we may
write an equation for the structure factor in the form: 
\begin{equation}
\partial _{t}S_{\vec{k}}=\sum_{\vec{k}^{\prime }}X_{\vec{k}\vec{k}^{\prime
}}S_{\vec{k}^{\prime }}\quad .  \label{S}
\end{equation}

An alternative approach starts with an equation of motion \cite{Newman}: $%
\partial _ts_{\vec{\imath}}(t)=\left[ s_{\vec{v}+\vec{b}}(t)-s_{\vec{v}%
}(t)\right] (\delta _{\vec{\imath}\vec{v}}-\delta _{\vec{\imath},\vec{v}+%
\vec{b}})$, where $\vec{v}$ is the position of the tag at time $t$, and $%
\vec{b}$ is a lattice vector which specifies the direction of the exchange. 
In Fourier space, $\tilde{s}_{\vec{k}}(t+1)=\sum_{\vec{k}%
^{\prime }}M_{\vec{k}\vec{k}^{\prime }}(\vec{v},\vec{b})\tilde{s}_{\vec{k}%
^{\prime }}(t)$, where $M_{\vec{k}\vec{k}^{\prime }}=\delta _{\vec{k}\vec{k}%
^{\prime }}-{\rm e}^{-i(\vec{k}-\vec{k}^{\prime })\vec{v}}(1-{\rm e}^{-i\vec{%
k}\vec{b}})(1-{\rm e}^{i\vec{k}^{\prime }\vec{b}})/L^d$. Since the
structure factor is defined by $S_{\vec{k}}(t)=\langle \tilde{s}_{\vec{k}}(t)%
\tilde{s}_{-\vec{k}}(t)\rangle ,$ where the average is over all possible
moves over the entire time period $\left[ 0,t\right] ,$ we
obtain, formally, 
\begin{multline}
S_{\vec{k}}(t) = 
\sum_{ \{\vec{p}_{i}\},\{\vec{q}_{i}\} }\big\langle
M_{\vec{k}\vec{p}_{t-1}}M_{-\vec{k}\vec{q}_{t-1}}
M_{\vec{p}_{t-1}\vec{p}_{t-2}}M_{\vec{q}_{t-1}\vec{q}_{t-2}}\cdots \\ \cdots
M_{\vec{p}_{1}\vec{p}_{0}} M_{\vec{q}_{1}\vec{q}_{0}}\big\rangle
\tilde{s}_{\vec{p}_0}(0)\tilde{s}_{\vec{q}_0}(0)\quad . 
\label{S(t)}
\end{multline}
For VMD, the allowed moves are correlated leading to a non-trivial average.
For KD, however, the position of the vacancy is randomized before each
exchange, so that the above average {\em factorizes} into $t$ factors. Cast
in the form of an evolution equation, the result is precisely Eq.~(\ref{S}),
with $X_{\vec{k}\vec{k}^{\prime }}=\left\langle M_{\vec{k}\vec{k}^{\prime
}}M_{-\vec{k}-\vec{k}^{\prime }}\right\rangle _{\vec{v},\vec{b}}-\delta _{%
\vec{k}\vec{k}^{\prime }}$ .

With either approach, thanks to the simplicity of $X$, the eigenvectors and
eigenvalues can be found exactly in the $L\rightarrow \infty $ limit.
Deferring details to \cite{ASZ_long}, we quote only the result for $d=2$.
With the reminder that $\vec{k}=(2\pi n/L,0)$ and the initial value
$S_{\vec{k}}(0)=\left( 2L/\pi n\right) ^{2}\delta _{n\text{ odd}}$, we obtain
\begin{multline}
S_{\vec{k}}(t)=S_{\vec{k}}(0)e^{-\left( 2\pi n/L^{2}\right) ^{2}t} \\
+\frac{4}{\pi^{2}}\sum_{l\neq n}\frac{n^{2}\left[ 1-e^{-\left( 2\pi
l/L^{2}\right) ^{2}t}\right] \delta _{l\text{ odd}}}{l^{2}(n^{2}-l^{2})}
\label{result}
\end{multline}

\begin{figure}[tbh]
\centerline{\epsfig{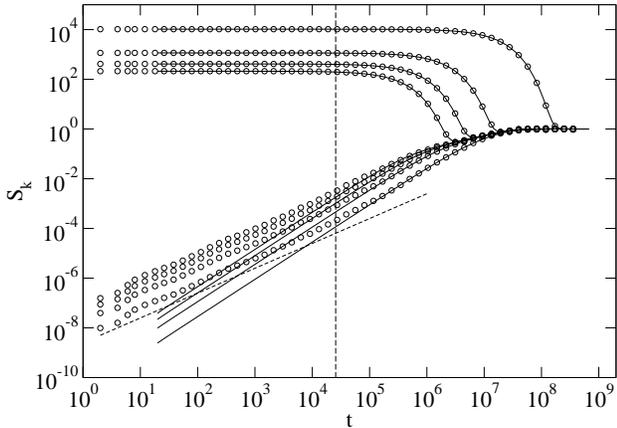}}
\caption{Comparison of Monte Carlo data (circles) for KD (same data as in
Fig.~\ref{fig:1}~(b)) and the exact calculation (solid lines) for the same
structure factors as in Fig.~\ref{fig:1}. The dashed line has slope 1, 
the solid lines have slope $3/2$ for small times.}
\label{fig:2}
\end{figure}

%Of course, the exclusion $l\neq n$ is unnecessary for even $n$'s. 
As in the ``two-fluid'' model, the two terms carry the same interpretations,
confirming a key mean-field result: the associated crossover time scaling with
$L^{4}$. In addition, comparing the first terms in
Eqs.~(\ref{eq:S(k)},\ref{result}), we may identify the diffusion constant $D$
as $1/\left( 2L^{2}\right) $. This value may be interpreted as the rate of
spin-exchange being equal to the hole density ($L^{-2}$) and the two
independent directions of its moves. The main improvement is the
$\vec{k}$-dependence in the disordered component. In more detail,
Eq.~(\ref{result}) correctly accounts for the presence (absence) 
of dips in the odd
structure factors for $n>1$ ($n=1$). Turning to early behavior
within the intermediate regime, we may extract remarkably detailed
information. For short times ($t\simeq L^{2}$), the behavior of the sum in
Eq.~(\ref{result}) is controlled by the large $l$ contributions, whence
$l^{2}-n^{2}\simeq l^{2}$. Using Poisson resummation, we find $S_{\vec{k}}(t)
\propto
n^{2}t^{3/2}$. Both the $t$- and the $n$-dependence are borne out by the
data. For larger times, $n^{2}t/L^{4}\simeq 1$, a third power law emerges
which is particularly noticeable for large $n$.  Here, the sum is dominated by
smaller $l$, so that $n^{2}-l^{2}\simeq n^{2}$. Resummation now yields
$t^{1/2}$, with an $n$-{\em independent} amplitude.  For large $n$, this power
law sets in at earlier times, so that structure factors with larger $n$ merge
{\em before} being joined by $S_{\vec{k}}$'s with smaller $n$.
Fig.~\ref{fig:2} shows a direct comparison of MC data with
Eq.~(\ref{result}). Excellent agreement is observed over many orders of
magnitude without {\em any} fit parameters.

\paragraph*{Conclusions.}

Using MC simulations and analytic arguments, we analyzed two
types of microscopic kinetics (vacancy-mediated and Kawasaki dynamics).
The structure factors display several remarkable features,
which characterize different temporal regimes. Just before the onset of
equilibration, the competition of surface- vs.\ bulk-dominated time scales
generates unexpected minima in several structure factors. These are
particularly prevalent in the ``odd'' structure factors that are sensitive to
the initial order. In contrast, the ``even'' structure factors grow via a
sequence of different power laws before saturating. This sequence includes a
$t^{3/2}$ regime for KD and a $t$ regime for VMD. Moreover, VMD generates dips
in even structure factors which are absent in KD. 
All of these features carry over to larger system sizes
and $d=3$; in fact, the range of the power law behaviors increases
significantly with system size. 
Thus, the structure factors
carry a clear signature of the underlying microscopic dynamics. We believe
that this could lead to a simple experimental identification of the dominant
kinetic mechanism in a material, e.g., via small-angle X-ray or neutron scattering.

Two complementary theoretical approaches -- a phenomenological coarse-grained
two-fluid model and an exact lattice calculation -- provide quantitative
insight into the physical origin of these features. 
Several questions remain open. First, our {\em exact} lattice analysis can
be carried out only for KD, where the dynamics of the spins is Markovian.
For VMD, in contrast, though the vacancy path is random, it induces
non-trivial correlations in the spins. Work is in progress to extend our
analysis to this case \cite{ASZ_long}. 
Presumably, these correlations
are also at the root of the observation that VMD equilibrates more slowly
than KD, by a factor of about 4.5 in $d=2$. It is not clear at this stage,
however, whether this this effect can be captured by an appropriate
generalization of correlation factors  \cite{corr-fac}. Finally,
the effect of several vacancies or tags should be investigated, as well as
upquenches to temperatures other than infinity. These cases have so far only
been studied at the mean-field level \cite{MFT}.

\paragraph*{Acknowledgements.}

We thank W.\ Triampo for discussions and A.\ Zippelius for hospitality at the
University of G{\"o}ttingen. This work is supported by the NSF under grant DMR-9727574 and 
the DFG under Zi209/5-1.

\end{document}